\documentclass{jfm}

\usepackage{float}
\usepackage{hyperref}
\usepackage{graphicx}
\usepackage{epstopdf}
\usepackage{amsmath}
\usepackage{amsfonts}
\usepackage{amssymb}
\usepackage{xcolor}
\usepackage{bm}
\usepackage{multirow}
\usepackage{natbib}

\newcommand{\re}{Re_\lambda}

\def\ff {{\mathbf{f}}}

\lefttitle{D. Buaria}
\righttitle{Scalar dissipation anomaly and scalar-gradient scaling
in turbulence}

\title{
Scalar dissipation anomaly and scalar-gradient scaling
in turbulence: A joint velocity-scalar
multifractal view
}

\author{Dhawal Buaria\aff{1,2}
}

\affiliation{
\aff{1}Department of Mechanical and Aerospace Engineering, Texas Tech University, Lubbock, TX 79409, USA
\aff{2}Max Planck Institute for Dynamics and Self-Organization, 37077 G\"ottingen, Germany
}

\corresau{Dhawal Buaria, \email{dhawal.buaria@ttu.edu}}

\begin{document}
\maketitle

\begin{abstract}

We revisit the problem of scalar dissipation anomaly 
and scaling of scalar gradients in passive scalar turbulence 
using theory and data from well-resolved 
direct numerical simulations (DNS) on grid sizes of 
up to $8192^3$, spanning Taylor-scale Reynolds
numbers $\re=140-1000$ and Schmidt numbers $Sc = 1-512$.  
The theory is based on a joint multifractal description 
of longitudinal velocity increments and scalar increments, 
constrained by Yaglom's law and extended to gradients via a 
fluctuating Batchelor cutoff scale.
The DNS data show that the normalized mean scalar dissipation 
approaches a single asymptotic value as both $\re$ and $Sc$ increase, 
although larger $Sc$ requires larger $\re$ to reach this state. 
In the multifractal framework, this corresponds to an effective scalar 
H\"older exponent tending to zero, associated with sharp cliff-like 
scalar fronts, and saturation of inertial-range scaling scalar 
structure-function exponents. The joint velocity-scalar fractal dimension of 
the dissipative structures is inferred to approach $7/3$, 
indicating a non-space-filling support. The framework further predicts that
for fixed $\re$, higher-order central moments of scalar gradients 
are independent of $Sc$. This prediction is confirmed 
by DNS data and by the collapse of standardized probability distributions 
of scalar-gradient across Schmidt numbers.  
These results suggest that the $Sc$-scaling of scalar gradients is 
dictated solely by scalar dissipation anomaly. In contrast, 
their $\re$-dependence reflects strong intermittency, which 
can be directly related to mixed velocity-scalar structure
function exponents.

\end{abstract}



\section{Introduction}

The transport and mixing of a passive scalar, such as temperature
or substance concentration, by a turbulent velocity field
governed by the Navier-Stokes equations is a canonical 
problem in fluid mechanics with broad relevance to 
atmospheric and oceanic transport,
combustion engineering, industrial mixing
and numerous other natural and technological flows.
From a phenomenological viewpoint,  
turbulent mixing proceeds through the continuous folding and 
fragmentation of 
scalar fluctuations by the velocity field, transferring 
scalar variance to progressively smaller scales until
the scalar diffusivity is strong enough
to dissipate them to complete mixing at the molecular level.
The key quantity characterizing this irreversible
destruction of scalar variance is the scalar dissipation rate
\begin{align}
\chi = 2 D \, \partial_i \theta \, \partial_i \theta \ , 
\label{eq:chi}
\end{align}
where $D$ is the scalar diffusivity and $\theta$ denotes the fluctuating
scalar field. 
Providing a direct measure of local mixing intensity, the
scalar dissipation rate plays a central role in theory and modeling
of turbulent mixing \citep{Sreeni97, Warhaft:2000, ShrSig00, Pitsch2000}. 

A fundamental question concerns the behavior of the mean scalar
dissipation $\langle \chi \rangle$ in the limit
of vanishing $D$, or equivalently at asymptotically large
Schmidt number $Sc \equiv \nu/D$, where $\nu$ is the kinematic
viscosity of the fluid. This is the scalar analogue
of the dissipative anomaly for turbulent kinetic energy.
For the energy dissipation rate
\begin{align}
\epsilon  = 2 \nu S_{ij} S_{ij}   \ , 
\qquad
S_{ij} = \tfrac{1}{2} (\partial_j u_i + \partial_i u_j) \ , 
\end{align}
where $u_i$ is the fluctuating velocity field,
it is well established that its mean $\langle \epsilon \rangle$
remains finite in the limit of decreasing $\nu$ despite the explicit
proportionality 
\citep{sreeni98, pearson02, Ishihara09, vassilicos15, Dubrulle:2019, buaria_jfmp}, 
or equivalently at large Reynolds number $Re = UL/\nu$, 
where $U$ and $L$ are characteristic large velocity and length scales.
This anomaly arises because, as viscosity decreases, 
the velocity field develops increasingly more intense gradients,
the two effects balancing such that $\langle \epsilon \rangle$
approaches a finite, viscosity independent limit. 
A directly analogous expectation arises for passive scalars:
as $D$ decreases, turbulent stirring generates increasingly intense
scalar gradients, such that $\langle \chi \rangle$ remains finite
in the asymptotic limit.

Scalar dissipation anomaly has been 
explored in a number studies over the years
\citep{borgas2004high, Donzis05, BYS.2016, BCSY2021b}.
The accumulated evidence from experiments
and direct numerical simulations (DNS)
firmly establishes its validity 
when $Sc=1$, provided that 
$Re$ is sufficiently high
\citep{Donzis05, BCSY2021b}.
However, the situation is markedly less clear in the regime
of simultaneously high Schmidt and Reynolds numbers.
As $Sc$ increases, scalar fluctuations extend
to scales increasingly smaller than the smallest scales of the
velocity field, imposing severe resolution
restrictions for DNS and experiments alike to 
accurately capture the scalar
gradients \citep{Batchelor1959}. Consequently, studies 
at high $Sc$ have often been restricted to low Reynolds numbers
when compared to $Sc\simeq1$
\citep{PK02, Donzis05, BCSY2021b}.
Such studies, predominantly utilizing DNS, have shown that when 
$Re$ is held fixed, the mean scalar dissipation
rate monotonically decreases with $Sc$, precluding
existence of an anomaly \citep{BCSY2021b}. 
However, whether such behavior persists at higher $Re$ 
still remains unresolved.

The issue is inseparable from the broader
problem of scaling of scalar gradients. 
It is well known that small scales and gradients 
(for both velocity and scalar) 
in turbulence exhibit strong intermittency 
\citep{Sreeni97, Warhaft:2000, PK02, BCSY2021a, BS2022}. 
While dissipative anomaly prescribes
the second moment of gradients --
forming the basis of classical mean-field phenomenologies 
\citep{K41a, Batchelor1959} --
intermittency necessitates anomalous scaling of higher order 
gradient moments \citep{Frisch95, SreeniYakhot:2021, buaria_jfmp}.
For velocity gradients, considerable progress has been 
made using the multifractal description. 
In this framework, inertial-range fluctuations are described by a 
spectrum of local H\"older exponents, which can then be connected to 
dissipative-scale gradients through a fluctuating viscous cutoff scale
\citep{Nelkin90, Paladin87, Frisch95}. 
This approach has provided a successful phenomenological description of 
anomalous velocity-gradient scaling \citep{buaria_jfmp}, 
and recent work has further extended 
it to reconcile the distinct scaling behaviors of longitudinal and 
transverse velocity gradients \citep{buaria2026}. 
For passive scalars, however, a comparable general framework is lacking. 
The scalar problem is intrinsically coupled: scalar gradients are produced 
by the straining and compressive motions of the velocity field
\citep{Ashurst87, Vedula:99, mishi26}, 
while the scalar increments themselves possess their 
own intermittent statistics \citep{Warhaft:2000}. 
A theory for scalar-gradient scaling must therefore account 
for the joint statistics of velocity and scalar fluctuations, 
rather than treating the scalar field in isolation.

In this work, we address these issues through a combined theoretical 
and numerical study of scalar dissipation anomaly and scalar-gradient scaling. 
The theoretical contribution is a joint multifractal framework 
for velocity and scalar statistics, obtained by extending the recent work of 
\citet{buaria2026} to passive scalar turbulence. 
The framework is formulated in terms of the coupled scaling of longitudinal velocity 
increments and scalar increments, motivated by Yaglom's exact relation for 
scalar variance cascade \citep{MY.II}. 
For $Sc>1$, its extension to gradients requires accounting for 
the fact that the scalar diffusive cutoff lies below the viscous cutoff 
of the velocity field. This naturally leads to a fluctuating Batchelor scale 
and connects scalar-gradient moments to the joint velocity-scalar multifractal spectrum. 
In this formulation, scalar dissipation anomaly is associated with the selection of scalar structures whose effective H\"older exponent $h_2^*$ tends to zero, 
corresponding to sharp cliff-like fronts in the scalar field,
also leading to saturation of scalar structure-function exponents. 

The theoretical predictions are assessed using high-resolution 
DNS data spanning Taylor-scale Reynolds numbers 
$\re=140-1000$ and Schmidt numbers $Sc=1-512$. 
The data show that scalar dissipation anomaly 
is realized at high $Sc$, but requires 
the Reynolds number to be significantly larger than 
that required for $Sc=1$. 
In the multifractal framework, the approach to the
asymptotic state is characterized by a weak power-law 
dependence of the form $Sc^{-h_2^*}$, with the data empirically
giving $h_2^* \approx 29.5/\re$, making this power-law
correction equivalent, at high $\re$, to the 
$\log Sc /\re$  correction 
derived previously \citep{borgas2004high, Donzis05}. 
We further show that, once the second moment is fixed by 
scalar dissipation anomaly, standardized higher-order 
scalar-gradient statistics become independent of $Sc$ at fixed $\re$; 
whereas their $\re$-dependence at fixed $Sc$ remains strongly intermittent 
and is tied to mixed velocity-scalar structure-function exponents. 
Together, these results provide a unified multifractal framework
for characterizing scalar dissipation anomaly and the
intermittency of scalar gradients in passive scalar turbulence.

The rest of the manuscript is organized as follows. 
In \S~2, we briefly describe the numerical approach 
and the DNS database. In \S~3, we develop the joint multifractal 
framework for velocity and scalar statistics. The DNS results are 
presented and analyzed in \S~4. Finally, \S~5 provides a 
summary of the main findings and concluding remarks.

\section{Numerical approach and DNS database}
\label{sec:setup}

We only briefly discuss the numerical approach 
as it has been 
described in several recent works 
\citep{BCSY2021b, BCSY2021a, BS2022, mishi26}. 
The DNS data utilized here correspond
to the canonical setup of 
forced stationary isotropic turbulence
in a triply periodic domain.
The velocity field is governed by 
the incompressible Navier-Stokes equations: 
\begin{align}
  \frac{\partial u_i}{\partial t} + u_j\frac{\partial u_i}
  {\partial x_j} &= -\frac{\partial P}{\partial x_i} + 
  \nu\nabla^2 u_i + f_i \ , \\ 
  \frac{\partial u_i}{\partial x_i} &= 0 \ ,
  \label{eq:NS}
\end{align}
where $P$ is the kinematic pressure and $\ff$ is the large-scale 
forcing term to maintain statistical stationarity 
\citep{EP88}.
The passive scalar field is obtained by solving 
the advection-diffusion equation in presence of a uniform
mean-gradient $\mathbf{G}$
\begin{align}
  \frac{\partial\theta}{\partial t} + u_j\frac{\partial
  \theta}{\partial x_j} = -G_j u_j + D\nabla^2\theta,
  \label{eq:theta}
\end{align}
imposed along the $x$-direction: $G_j = G \delta_{1j}$,
providing the forcing to maintain the scalar field in a statistically
stationary state \citep{pumir94, Overholt1996}.

\begin{table}
\centering
\setlength{\tabcolsep}{10pt}
\begin{tabular}{ccccccc}
$\re$ & $Sc$ & $N_u^3$ &  $k_{\max}\eta$ & $N_\theta^3$ & $k_{\max}\eta_B$ 
& $T_{\mathrm{sim}}/T_E$ \\
\hline
\multirow{8}{*}{140}
 & 1    & $1024^3$  & 6 & $1024^3$ & 6 & 10 \\
 & 8    & $1024^3$  & 6 & $1024^3$ & 2 & 10 \\
 & 16   & $1024^3$  & 6 & $2048^3$ & 3 & 15 \\
 & 32   & $1024^3$  & 6 & $2048^3$ & 2 & 11 \\
 & 64   & $1024^3$  & 6 & $4096^3$ & 3 & 9 \\
 & 128  & $1024^3$  & 6 & $4096^3$ & 2 & 12 \\
 & 256  & $1024^3$  & 6 & $8192^3$ & 3 & 6 \\
 & 512  & $1024^3$  & 6 & $8192^3$ & 2 & 9 \\
\hline
\multirow{2}{*}{240}
 & 1  & $2048^3$  & 6 & $2048^3$ & 6 & 8 \\
 & 8 & $2048^3$  & 6 & $2048^3$ & 2 & 8 \\
\hline
\multirow{2}{*}{390}
 & 1 & $4096^3$  & 6 & $4096^3$ & 6 & 3 \\
 & 8  & $2048^3$  & 6 & $8192^3$ & 4 & 6 \\
\hline
\multirow{2}{*}{650}
 & 1 & $6144^3$  & 4.5 & $6144^3$ &  4.5 & 3 \\
 & 8 & $8192^3$  & 6   & $8192^3$ &   2  & 2 \\
\hline
1000 & 1  & $8192^3$ & 3 & $8192^3$ & 3 & 2 \\
\hline
\end{tabular}
\caption{
Simulation parameters for the DNS runs used in the current 
work: the Taylor-scale Reynolds number $\re$, the Schmidt number $Sc$, 
the number of grid points for the velocity and scalar fields $N_v^3$ 
and $N_\theta^3$, the spatial resolution for the velocity and scalar 
fields $k_{\max}\eta$ and $k_{\max}\eta_B$, and the simulation length 
$T_{\mathrm{sim}}/T_E$ in statistically stationary state in terms of 
the large-eddy turnover time $T_E$. Cases with $N_\theta = N_u$ are 
solved using conventional pseudospectral method for both velocity and
scalar fields, while cases with 
$N_\theta > N_u$ employ a hybrid spectral-compact method
\citep{gotoh12a, clay.cpc1, clay.cpc2}.
}
\label{tab:dns}
\end{table}

The full set of DNS database is given in 
Table~\ref{tab:dns}, and combines all the runs
in \citet{BCSY2021b, mishi26}.
As also outlined in \citet{mishi26},
for $Sc=1,8$, we utilize conventional 
Fourier pseudospectral methods to solve both 
velocity and scalar fields, with aliasing
errors controlled via a combination of truncation 
and phase shifting  \citep{Rogallo, PattOrs71}. 
For $Sc>8$, a hybrid approach is used 
\citep{gotoh12a, clay.cpc1, clay.cpc2},
whereby the velocity field is solved
pseudospectrally by appropriately
resolving the Kolmogorov scale $\eta_K$ but the scalar field
is solved using high-order compact finite schemes
on a finer grid to resolve the 
Batchelor scale $\eta_B = \eta_K \, Sc^{-1/2}$ 
\citep{Batchelor1959}.
While many runs are same as the ones utilized
in \citet{BCSY2021a, BCSY2021b}, 
we have performed newer
runs at higher Reynolds numbers \citep{mishi26}.

\section{Joint multifractal description of velocity and scalar intermittency}

Multifractal descriptions have long provided a useful 
phenomenological framework for turbulence intermittency, 
especially for characterizing anomalous scaling of velocity 
increments and gradients; see e.g., 
\citet{benzi1984, MS91, Frisch95, Sreeni97, Dubrulle:2019, buaria_jfmp} 
and references therein. However, such descriptions have overwhelmingly
been univariate -- they describe 
the statistics of a single fluctuating quantity, typically the
longitudinal velocity increment, through a spectrum of local scaling 
exponents. By contrast, joint or multivariate 
multifractal descriptions have been 
surprisingly rare \citep{meneveau1990joint, buaria2026}, 
despite the fact that turbulence is 
inherently a problem of coupled fluctuating fields 
and tensorial quantities.

In an early work, \citet{meneveau1990joint} motivated
joint descriptions for turbulence intermittency
and used them to characterize
inertial range scaling of locally averaged
quantities. 
More recently, \citet{buaria2026} considered a 
joint framework
explicitly for longitudinal and transverse  
velocity increments, and also gradients. 
That work showed that a bivariate multifractal description 
can capture information that is fundamentally inaccessible by 
separate univariate descriptions, particularly when the quantities of 
interest are dynamically coupled but possess distinct intermittent statistics.

The extension of this idea to passive scalar turbulence is natural. 
Scalar mixing involves the coupled action of velocity and scalar fluctuations
to drive the scalar variance cascade from large to diffusive scales. 
A general joint multifractal description could therefore be formulated 
in terms of three fluctuating quantities: the longitudinal velocity increment, 
the transverse velocity increment, and the scalar increment. 
However, here we will consider a bivariate 
description involving only the longitudinal velocity increment and
and the scalar increment. 
This choice is for two important reasons.

First, any reliable multifractal description must be consistent 
with the exact third-order inertial-range relations derivable
from the governing equations. For the velocity field,
the $4/5$-th and $4/15$-th laws \citep{eyink2003local}
provide such exact relations
that motivate the need for both longitudinal and transverse 
velocity increments when constructing a 
joint multifractal description
for the velocity field alone \citep{buaria2026}. 
For passive scalar turbulence,
the corresponding exact result is Yaglom's law \citep{MY.II}, 
which couples the longitudinal velocity and scalar increments,
obviating the need for transverse increments. 
Second, a recent analysis of 
amplification mechanisms for scalar gradients \citep{mishi26}
shows that it is overwhelmingly driven by the strain-rate, 
i.e., by longitudinal gradients, with
transverse gradients playing an insignificant role.
Consistent with expectation from Yaglom's
law, this provides another physical reason
to focus only on a joint description of longitudinal
velocity increments and scalar increments.

\subsection{Joint description of increments}

We  consider the increments
\begin{align}
 \delta u_r = u(x+r) - u(x)  \ , \qquad 
 \delta \theta_r = \theta(x+r) - \theta(x) \ , 
\end{align}
where $\delta u_r$ is the longitudinal velocity increment 
and $\delta \theta_r$ is the scalar increment,
across a separation $r$ along the $x$-direction. 
The central idea in the multifractal formulation
is that these increments are H\"older continuous
\begin{align}
\delta u_r \sim U \, (r/L)^{h_1}  \ ,  \qquad
\delta \theta_r \sim \Theta \, (r/L)^{h_2}  \ ,
\end{align}
with the exponents $h_1$ and $h_2$ realized
on a set with fractal dimension  $D(h_1, h_2)$, which
is also termed the joint multifractal spectrum.
Here, $U$ and $\Theta$ correspond to rms of velocity and
scalar fluctuations, respectively, and $L$ is the large-eddy
length scale. 
The joint spectrum implies that the probability
to observe the exponents $h_1, h_2$ at scale-size $r$
is given as:
\begin{align}
P_r (h_1, h_2) \sim  (r/L)^{3 - D(h_1, h_2)}    
\end{align}
With these definitions, the mixed velocity-scalar
structure function of order $p_1, p_2$ can be obtained
by integrating over full bivariate distribution
\begin{align}
\langle (\delta u_r)^{p_1} \,
(\delta \theta_r)^{p_2} \rangle \sim  U^{p_1} \, \Theta^{p_2} \,
\int \int  \left( \frac{r}{L} \right)^{p_1 h_1 + p_2 h_2 + 3 - D(h_1, h_2)} \ dh_1 dh_2  \ . 
\end{align}
Thereafter, the steepest descent estimation in the limit $r/L\to0$
leads to the result
\begin{align}
\langle (\delta u_r)^{p_1} \,
(\delta \theta_r)^{p_2} \rangle \sim  U^{p_1} \, \Theta^{p_2} \,
\left( \frac{r}{L} \right)^{\zeta_{p_1,p_2}}  \ , 
\end{align}
with the scaling exponents given as
\begin{align}
\zeta_{p_1, p_2} &= \inf_{h_1, h_2} \,
[ p_1 h_1 + p_2 h_2 + 3 - D(h_1, h_2) ]  \ .
\label{eq:zetap}
\end{align}
Note that the infimum essentially implies that 
for a given $p_1, p_2$ value, the critical 
H\"older exponents which dictate the scaling exponent 
are obtained by solving the system: 
\begin{align}
\frac{\partial D}{\partial h_1} (h_1^*, h_2^*) = p_1  \ , 
\qquad
\frac{\partial D}{\partial h_2} (h_1^*, h_2^*) = p_2  \ .
\label{eq:hcrit}
\end{align}

The univariate velocity and scalar exponents
are recovered as special cases
for $p_2=0$ and $p_1=0$, respectively. However,
the joint information 
provided by $p_1 , p_2 \neq 0$ cannot be in general inferred
from the two marginal spectra alone. This is evident when 
considering Yaglom's law
\begin{align}
\langle (\delta u_r) \,
(\delta \theta_r)^{2} \rangle = -\frac{4}{3} \, \langle\chi\rangle \, r \ .
\label{eq:yaglom}
\end{align}
which can be exactly derived from the governing equations. 
It follows for this case that $\zeta_{1,2}=1$, 
and thus
\begin{align}
h_1^* + 2 h_2^* + 3 - D(h_1^*, h_2^*) = 1  \ .
\label{eq:yaglom_dh}
\end{align}
where $h_1^*$ and $h_2^*$ correspond to 
critical exponents for the minimization
procedure in Eq.~\eqref{eq:zetap}
when $p_1=1$ and $p_2=2$. 
Note that similar considerations apply for
joint description of longitudinal and transverse velocity increments,
with the Yaglom's law replaced by the $4/15$-th law derived
directly from Navier-Stokes equations \citep{buaria2026}. 


\subsection{Extension to gradients}

Defining the velocity and scalar gradients
as $\partial_x u \sim \delta u_r/r$ 
and  $\partial_x \theta \sim \delta \theta_r/r$,
respectively, we can write
\begin{align}
\partial_x u \sim  \frac{U}{L} \left( \frac{r}{L} \right)^{h_1 -1} \ \Bigg|_{r=\eta_u} \ ,
\ \ \ \ \ \ \ 
\partial_x \theta \sim  \frac{U}{L} \left( \frac{r}{L} \right)^{h_2 -1} \ \Bigg|_{r=\eta_\theta}  \ , 
\label{eq:jgrad}
\end{align}
where $\eta_u$ and $\eta_\theta$ are the dissipative cutoff scales
for velocity and scalar fields, respectively. 
This is where the joint framework for velocity and scalar statistics 
becomes fundamentally different than that of longitudinal 
and transverse velocity statistics considered in 
\citet{buaria2026}. For the latter, both quantities being components
of the same velocity field, are regularized by the same cutoff scale $\eta_u$.
In contrast, the dissipative cutoff for scalar field $\eta_\theta$
is different when $Sc\neq1$. 

For the velocity field, the local dissipative cutoff is 
obtained in the usual way by requiring the scale-dependent 
Reynolds number to be of order unity
\begin{align}
\frac{\delta u_r \  r}{\nu} \, \Big|_{r = \eta_u} \simeq  1   \ .
\label{eq:re1}
\end{align}
leading to the known result
\begin{align}
\eta_u/L \sim Re^{-1/(1 + h_1)} \ .
\label{eq:etah}
\end{align}
It is easy to see that the Kolmogorov
length scale $\eta_K = \eta_u (h_1 = \tfrac{1}{3})$. 

For the scalar field, one might generalize 
Eq.~\eqref{eq:re1} by replacing $\nu$ with $D$, giving
\begin{align}
\frac{\delta u_r \  r}{D} \, \Big|_{r = \eta_\theta}= 1   \ .
\label{eq:re1_sc}
\end{align}
leading to the result
\begin{align}
\eta_\theta/L \sim Re^{-1/(1 + h_1)} \, Sc^{-1/(1 + h_1)} \ .
\label{eq:etah_sc}
\end{align}
However, some care must be taken in interpreting and using this result.
The above result is appropriate when the scalar cutoff lies within
the range over which the velocity increment is still H\"older
continuous. This would be relevant for  
$Sc \leq 1$, for which  $\eta_u \leq \eta_\theta$.
In fact, it can be seen that  $h_1 = \frac{1}{3}$
gives the result 
$\eta_\theta = \eta_u Sc^{-3/4}$ which corresponds to 
the Obhukov-Corrsin scale \citep{MY.II}. 

However, here our focus is $Sc>1$, for which the situation is different. 
In this regime,
$\eta_\theta < \eta_u$, and scalar fluctuations persist
below the viscous cutoff scale of the velocity field.  
The scalar field is therefore advected 
by a smooth velocity field, as opposed to a rough one, i.e.,
the velocity field can no longer be considered H\"older continuous. 
Thus, for $Sc>1$, the correct estimate for scalar cutoff scale 
corresponds to $h_1 = 1$ for the sub-viscous part of the scaling,
which shows up in the $Sc$ dependence. 
This leads to the result
\begin{align}
\eta_\theta (h_1) = \eta_u (h_1)  \, Sc^{-1/2} \ ,  
\label{eq:etah_sc12}
\end{align}
which is essentially a fluctuating version of the Batchelor scale,
with $h_1 =\tfrac{1}{3}$ giving back the classical estimate 
$\eta_B = \eta_K Sc^{-1/2}$ \citep{Batchelor1959}. 

We can now write an expression for 
the joint moment of longitudinal velocity and scalar gradients, of
orders $n_1$ and $n_2$, respectively
\begin{align}
&\langle (\partial_x u)^{n_1} 
(\partial_x \theta )^{n_2} \rangle \nonumber \\ 
&\sim  
\left(\frac{U}{L}\right)^{n_1} \,
\left(\frac{\Theta}{L}\right)^{n_2}
\int \int \left( \frac{r}{L} \right)^{n_1(h_1-1) + n_2(h_2-1) + 3 - D(h_1, h_2)}  
\, Sc^{\frac{n_2 (1-h_2)}{2}} \, dh_1 \, dh_2  \ \Bigg|_{r=\eta_u}    
\end{align} 
Substituting $\eta_u/L$ from Eq.~\eqref{eq:etah} and
using steepest descent estimate for $Re \to \infty$, it follows
that
\begin{align}
\langle (\partial_x u)^{n_1} 
(\partial_x \theta )^{n_2} \rangle \sim  
\left(\frac{U}{L}\right)^{n_1} \,
\left(\frac{\Theta}{L}\right)^{n_2} \ 
Re^{\rho_{n_1,n_2}}  \ Sc^{\sigma_{n_2}}
\label{eq:derv_n1n2}
\end{align}
where the scaling exponents are given as  
\begin{align}
\rho_{n_1,n_2} &= \sup_{h_1,h_2}  \ \frac{n_1(1 - h_1) + n_2(1 - h_2) - 3 + D(h_1, h_2)}{1+h_1}  \ .
\label{eq:rhon} 
\end{align}
and
\begin{align}
\sigma_{n_2} &=  \frac{n_2}{2}(1 - h_2^*) 
\end{align}
where $h_2^*$ corresponds to the critical exponent 
for obtaining  $\rho_{n_1, n_2} $
in Eq.~\eqref{eq:rhon}, for a chosen value of $n_1, n_2$. 

Since the same joint multifractal spectrum
governs both the inertial-range exponents $\zeta_{p_1, p_2}$ 
and the gradient exponents $\rho_{n_1, n_2}$, 
it is easy to show that knowledge
of one also prescribes the other.
For instance, for a chosen $n_1, n_2$, the saddle point
$(h_1^*, h_2^*)$ that determines 
$\rho_{n_1, n_2}$, also determines
and inertial-range mixed structure function exponent
for some pair $(p_1, p_2)$. As derived in \citet{buaria2026}, 
one obtains 
\begin{align}
 \rho_{n_1, n_2} = p_1 (n_1, n_2) - n_1 \ ,   
\end{align}
where $p_1 (n_1, n_2)$ is a solution to the system
\begin{align}
\label{eq:p_def_1}
\zeta_{p_1, p_2} + (p_1 + p_2) &= 2 (n_1 + n_2)  \ , \\
p_2 &= n_2 \ .
\label{eq:p_def_2}
\end{align}
In other words, the gradient scaling exponents may be obtained
from inertial-range mixed exponents without the explicit
knowledge of the multifractal spectrum. 

Before we utilize the above framework for scalar gradient
statistics, it is useful to briefly discuss the limiting case $n_2=0$
(and thus $p_2 = 0$),
corresponding solely to scaling of longitudinal velocity gradients.
In this case, 
Eq.~\eqref{eq:p_def_1} leads to
\begin{align}
\zeta_{p_1, 0} + p_1 = 2 n_1    
\end{align}
with $\rho_{n_1, 0} = p_1 (n_1) - n_1$,  
and additionally $\sigma_{n_2} = 0$. 
This is precisely the usual univariate multifractal
result for longitudinal statistics
\citep{Nelkin90, Frisch95}.
In fact, this reduction is to be expected, since a passive
scalar cannot influence the scaling of the velocity field. 
However, the converse is not true and scalar gradient statistics
are indeed affected by the velocity field.

\subsection{Scaling of scalar gradients}
\label{subsec:scgrad}

Let us now consider the case $n_1 = 0$ and $n_2 = n$
(and $p_2 = n$), 
which corresponds to the 
$n$-th order moment of scalar gradients. 
In this case, Eq.~\eqref{eq:p_def_1}
becomes 
\begin{align}
\zeta_{p-n,n} + p &= 2n  \ , \qquad {\rm with } \quad p = p_1 + p_2 \ , 
\label{eq:zeta_t}
\end{align}
which shows that scaling of scalar gradients
is controlled not by scalar structure functions alone,
but by mixed velocity-scalar structure
functions. This result is directly analogous to that of \citet{buaria2026},
where transverse velocity gradient statistics were shown 
to be controlled by mixed longitudinal-transverse velocity 
structure functions. 
Here, the role of the transverse increment is played by the scalar increment, 
while the longitudinal velocity controls the 
local advective transfer of scalar variance.

The above result is also physically consistent with 
Yaglom's law given earlier in Eq.~\eqref{eq:yaglom} 
and scalar dissipation
anomaly. 
For $n=2$, it can be seen  Eq.~\eqref{eq:zeta_t} is 
exactly satisfied for $p_1, p_2 = (1,2)$ 
and $\zeta_{1,2}=1$, which is precisely the Yaglom's law. 
This leads to the result $\rho_{0,2} = p_1 - n_1 = 1 - 0 =  1$,
and additionally $\sigma_2 = 1 - h_2^*$. 
Thus, we can write the result in Eq.~\eqref{eq:derv_n1n2} as:
\begin{align}
\langle 
(\partial_x \theta )^2 \rangle \sim  
\frac{\Theta^2}{L^2}  \ 
Re^{1}  \ Sc^{1-h_2^*} \ . 
\end{align}
Using $Re=UL/\nu$ and $Sc=\nu/D$, and rearranging, the above gives
\begin{align}
\langle \chi\rangle \sim D \langle (\partial_x \theta )^2 \rangle \sim  
\frac{\Theta^2 \, U}{L}  \ Sc^{-h_2^*}
\label{eq:chi_h2}
\end{align}
which provides the multifractal prediction for scaling
of mean scalar dissipation rate. 

It can be seen right away that the above result suggests
that scalar dissipation is independent of Reynolds number. 
But otherwise, a residual dependence on $Sc$ remains through
the scalar H\"older exponent $h_2^*$. Thus, a true scalar 
dissipation anomaly is realized if $h_2^* = 0$ or at least
asymptotically approaches zero. If $h_2^* > 0$, the result
predicts that $\langle \chi \rangle$ decreases with $Sc$.
Remarkably, this indeed has been the observation
from prior results, see e.g. \citet{Donzis05, BCSY2021b}. 
We will revisit them soon in the next section, together with 
new DNS data and demonstrate that they are near-perfectly consistent
with the predictions above. 

It is worth recalling the following result for 
scaling of scalar dissipation rate,
derived earlier by \citet{borgas2004high}: 
\begin{align}
\frac{\Theta^2 U}{\langle\chi\rangle L} =  c_1  + c_2 \, \frac{\log Sc}{Re_\lambda} 
\label{eq:chi_borg}
\end{align}
where $c_1$ and $c_2$ are some constants. 
Its derivation relies on a piecewise integration of scalar
spectrum by assuming Obhukov-Corrsin's $k^{-5/3}$ scaling in the 
inertial-convective range and Batchelor's $k^{-1}$ scaling
in the viscous-diffusive range. Since these scalings are only
nominally satisfied at very high $Re$ and $Sc$, and even otherwise 
there exists transition regimes between these scalings, 
the result in Eq.~\eqref{eq:chi_borg} should be
understood as a semi-empirical one. 
It essentially suggests that scalar dissipation has a weak $\log Sc$
dependence, which diminishes with increasing Reynolds number, with a true
anomaly only realized when $(\log Sc)/Re_\lambda \ll 1$. 

Interestingly, this result can be reconciled with our own 
in Eq.~\eqref{eq:chi_h2}, by noting
$Sc^{h_2^*} = \exp ( h_2^* \log Sc)$
and realizing that if $h_2^*$ is small, the exponential term
can be approximated as
\begin{align}
\exp ( h_2^* \log Sc) \approx 1 + h_2^* \log Sc  \  , 
\qquad  {\rm for} \quad h_2^* \to 0  \ . 
\end{align}
Thus, in the limit
of small $h_2^*$ it follows that
\begin{align}
\frac{\Theta^2 U}{\langle\chi\rangle L} \sim 1 + h_2^* \log Sc    
\end{align}
which can be exactly mapped to Eq.~\eqref{eq:chi_borg} for 
\begin{align}
h_2^* \approx (c_2/c_1)  \, \frac{1}{Re_\lambda}  
\label{eq:h2_re}
\end{align}
Thus, while the semi-empirical analysis of \citep{borgas2004high} 
predicts a $\log Sc$ dependence, it can be reconciled
with our multifractal result of a weak power-law $Sc$-dependence,
with the  additional expectation
that the exponent $h_2^*$ further decreases
with Reynolds number. 
This is to essentially say that $h_2^*$ asymptotically
approaches zero at high $Re$, recovering a true anomaly
where scalar dissipation rate is independent of both
$Re$ and $Sc$.





\section{Results}

\subsection{Scalar dissipation anomaly}

\begin{figure}
\centering
\includegraphics[height=0.38\linewidth]{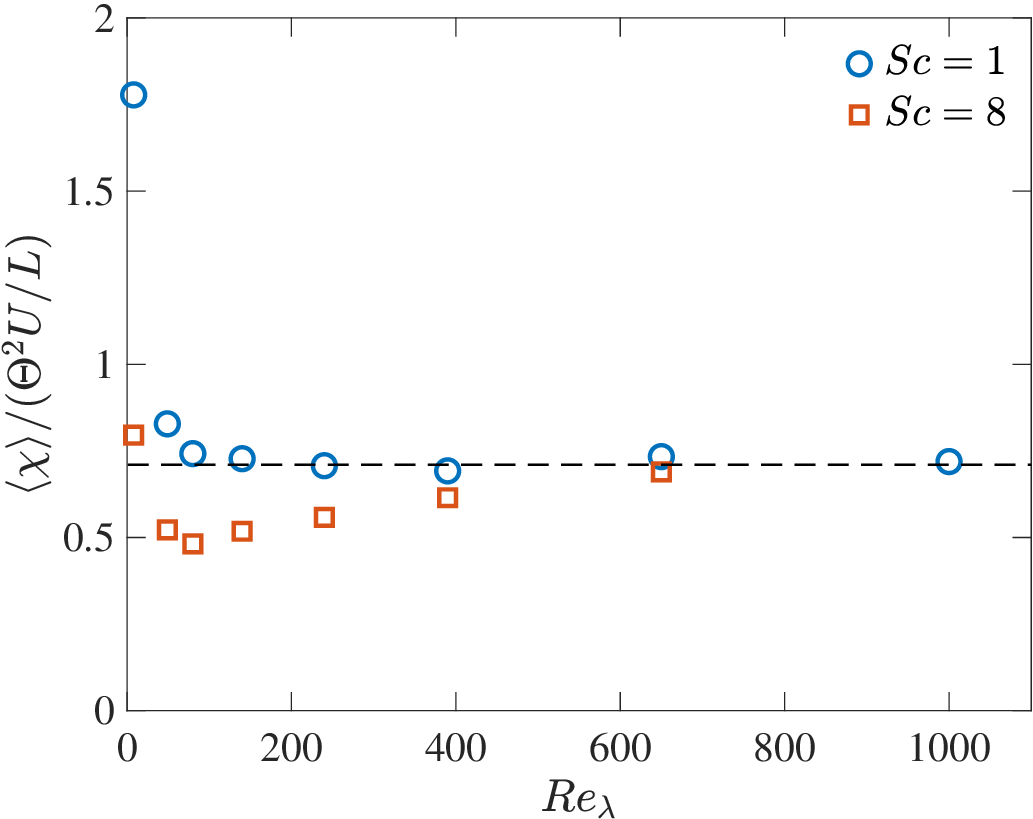} \ \ \ \ \ 
\includegraphics[height=0.38\linewidth]{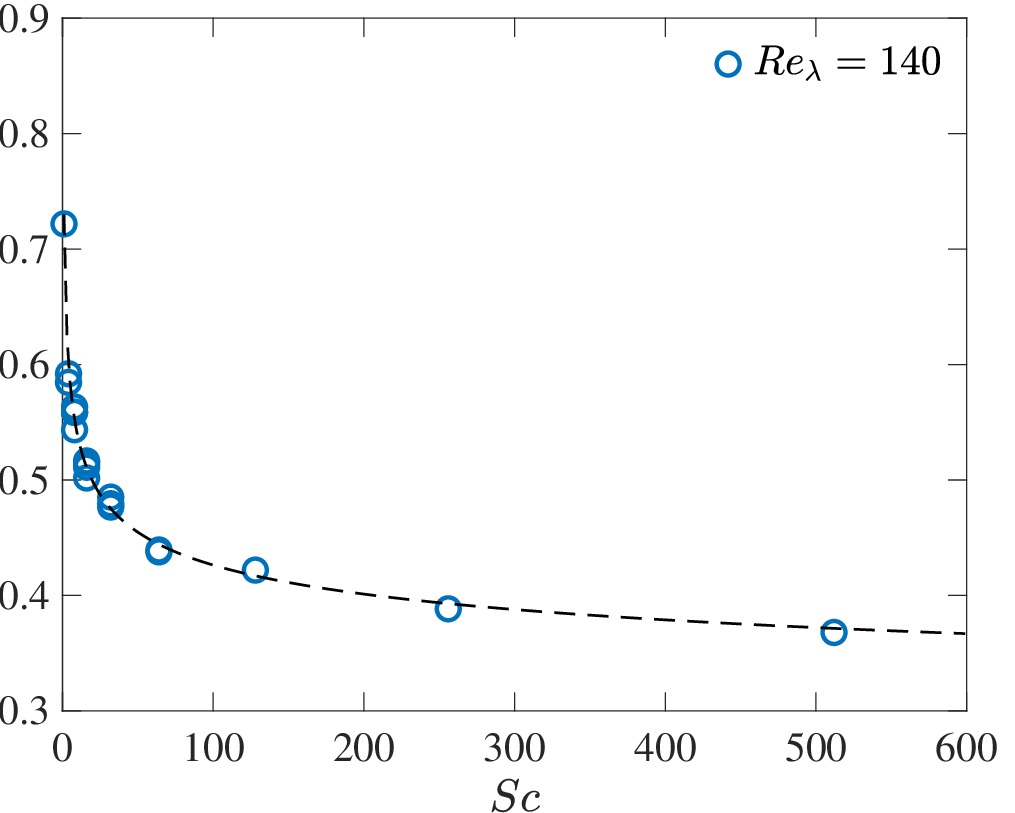} 
\caption{
Normalized mean scalar dissipation rate 
as a function of:  a) $\re$ for $Sc=1$ and $8$, 
and b) $Sc$ for $\re=140$. 
}
\label{fig:scds1}
\end{figure}

We now test the predictions of the preceding section 
using the DNS data.
Figure~\ref{fig:scds1}a shows the
dependence of normalized $\langle \chi\rangle$
on $Re_\lambda$ 
for $Sc=1,8$. For $Sc=1$, it can be readily
seen that the normalized $\langle \chi\rangle$ asymptotes
to a constant value  of $\approx 0.7$ (horizontal dashed line) 
for $\re \gtrsim 140$. 
This behavior is consistent with previous results
of \citet{Donzis05, BCSY2021b}, 
and also extends it to higher Reynolds number of 
up to $\re=1000$.
In contrast, the behavior for $Sc=8$ is markedly different.
At lower $\re$, the normalized  $\langle \chi\rangle$
remains significantly below the $0.7$ value, indicating a lack
of anomaly at $\re\gtrsim140$. 
However, with increasing $\re$, the data points
gradually approach the value of $0.7$, suggesting
that anomaly is indeed recovered. 
This observation was not realized in earlier works,
as the DNS runs at $Sc=8$ were restricted to much 
lower $\re$. 

The complementary dependence of the normalized 
$\langle \chi\rangle$
on $Sc$ is shown in Fig.~\ref{fig:scds1}b, 
at fixed $\re=140$. 
Although this result was shown previously 
in \citet{BCSY2021b}, we include it here 
to draw an explicit comparison with the result in Fig.~\ref{fig:scds1}a.
In this case, the mean scalar dissipation rate
monotonically decreases with increasing $Sc$,
as also expected from the theoretical predictions.
Thus, the simple picture that emerges 
is that scalar dissipation anomaly
is realized for any $Sc$ as the Reynolds number 
increases, with higher $Sc$ requiring higher Reynolds number to reach
the asymptotic state. However, if the Reynolds number is fixed, arbitrarily 
increasing the $Sc$ still violates the anomaly. 
This distinction emphasizes that scalar dissipation anomaly
is controlled not by $Sc$ alone, but rather the ability of turbulent
velocity field to generate sufficiently singular scalar gradients,
in the limit of vanishing viscosity.
This view is also consistent with the broader interpretation of turbulent
scalar transport in zero-diffusivity limit, where the loss
of smoothness of Lagrangian trajectories and onset of spontaneous
stochasticity provide a mechanism for irreversible scalar mixing
\citep{falkovich01, drivas2017}.

\begin{figure}
\centering
\includegraphics[width=0.48\linewidth]{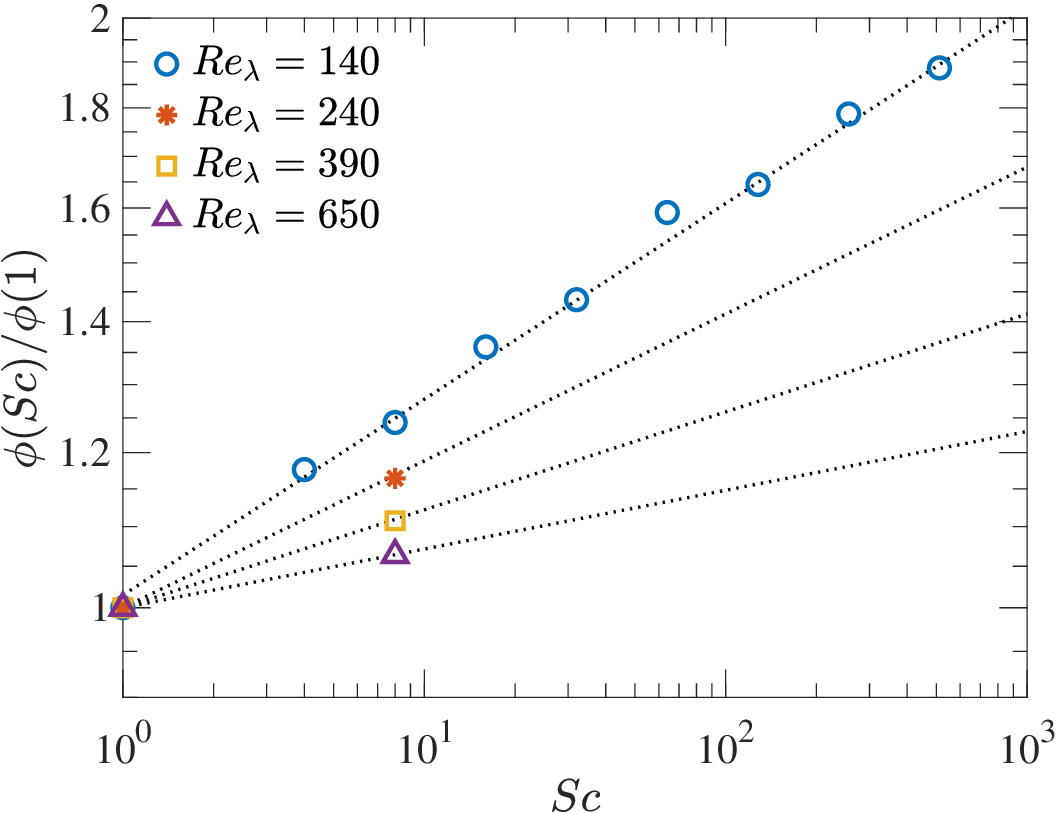} \ \ \
\includegraphics[width=0.48\linewidth]{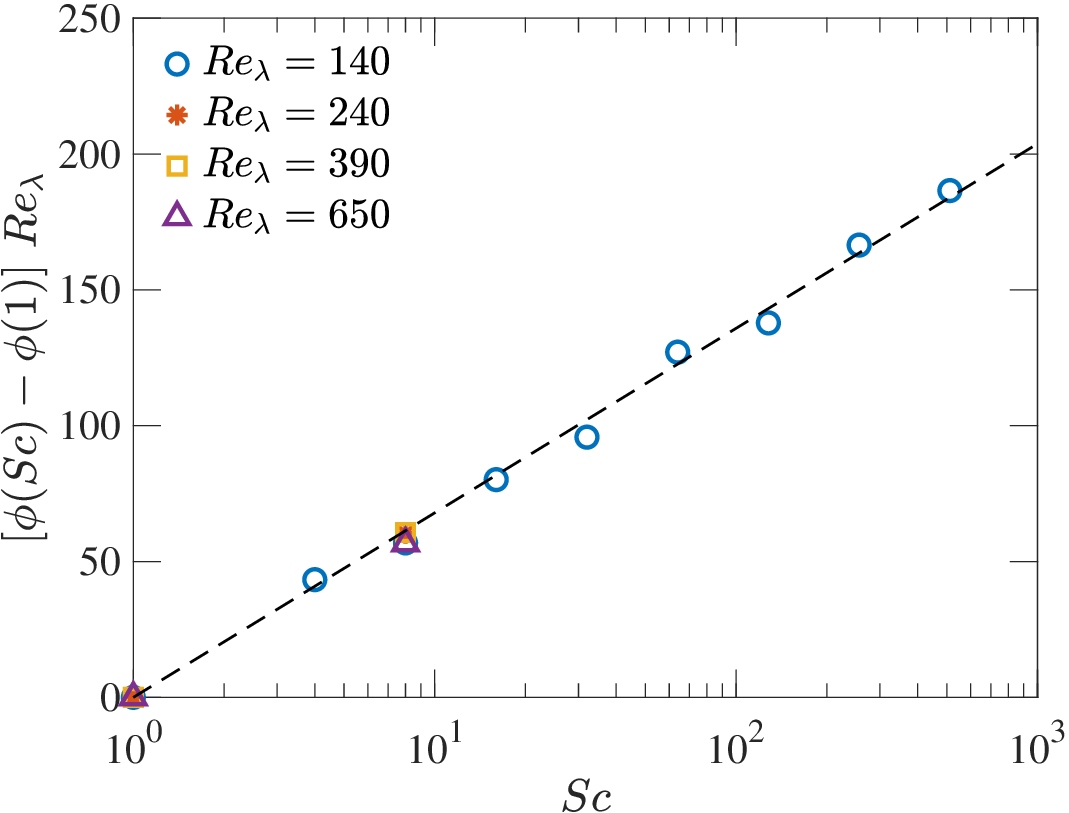}
\caption{
Schmidt number dependence of the quantity $\phi$,
the reciprocal of normalized mean scalar 
dissipation rate as defined by Eq.~\eqref{eq:phi}, at various $\re$. 
Panel a tests the power-law dependence given in Eq.~\eqref{eq:chi_h2},
using log-log scale.
Panel b tests the logarithmic dependence given in 
Eq.~\eqref{eq:chi_borg}, using log-linear scale.
}
\label{fig:scds2}
\end{figure}

To precisely quantify the behavior of $\langle \chi \rangle$ 
and connect it to earlier predictions, we define
the quantity
\begin{align}
 \phi = \frac{\Theta^2 U}{\langle \chi \rangle L}  \ ,
 \label{eq:phi}
\end{align}
which is essentially the inverse of normalized scalar dissipation rate.  
As discussed in \S~\ref{subsec:scgrad}, the scaling of this quantity
can be characterized in two complementary ways. The multifractal
result in Eq.~\eqref{eq:chi_h2} gives a power-law $Sc^{h_2^*}$ dependence,
whereas the alternative result in Eq.~\eqref{eq:chi_borg} gives a 
$\log Sc$ dependence.
These two results can be reconciled through a Reynolds number dependence
of the critical H\"older exponent $h_2^*$, which decreases was the 
Reynolds number increases. 
To first test the power-law prediction, 
Fig.~\ref{fig:scds2}a shows $\phi$ as a function of $Sc$ for different
$\re$, divided by its value at $Sc=1$ to account for any minor
statistical variations in the asymptotic value.  
For each $\re$, the data are perfectly  
consistent with a power-law dependence on $Sc$, with the exponent systematically
decreasing as $\re$ increases. 
At $\re=140$, the effective exponent is $h_2^*\approx 0.1$, consistent with
previous analysis of \citet{BCSY2021a}.
At $\re=650$, the exponent decreases to $h_2^* \approx 0.03$.
Extrapolating this to higher $\re$ leads to the expectation
$h_2^* \to 0$, thereby recovering the expected anomaly. 
A simple fit over the observed trend 
gives the result $h_2^* \approx 20/\re$.

We next test the log-dependence predicted by Eq.~\eqref{eq:chi_borg}.
Figure~\ref{fig:scds2}b 
shows the quantity $(\phi - \phi_{Sc=1})$, multiplied
by $\re$, as a function of $Sc$ on a log-linear scale -- 
the expectation being that this quantity
scales as $c_2 \log Sc$, with the same $c_2$ for all $\re$.
Indeed, the data show a striking collapse onto a single
curve with $c_2 \approx 29.5$.
Incidentally, \citet{borgas2004high} also provided a prediction
for this constant: $c_2 = \tfrac{ 5 \sqrt{15}}{3} B_\theta$,
where $B_\theta$ is the Batchelor constant corresponding to the
$k^{-1}$ scaling in the viscous-diffusive range. 
Using the known estimate $B_\theta \approx 5$ \citep{donzis2010batchelor}
gives $c_2 \approx {32.3}$, which is 
reasonably close to the observed value of $29.5$, especially 
given the semi-empirical nature of the derivation in \citet{borgas2004high}.

\subsection{Significance of the exponent $h_2^*$}
\label{subsec:h2}

The analysis so far leads to a natural question 
regarding the significance and physical implications 
of the critical exponent $h_2^*$.  For a strict
scalar dissipation anomaly with respect to both Reynolds
and Schmidt numbers, $h_2^*$ must be zero.
This is indeed supported by DNS data. The correspondence 
between the logarithmic correction and the multifractal result
in Fig.~\ref{fig:scds2} gives $h_2^* \approx 20/\re$.
This leads to $h_2 \approx 0.03$ at $\re=650$, which is already 
quite close to the asymptotic limit of zero. 
Since the exponent $h_2$ dictates
the roughness of scalar increments, $h_2 \to 0$ 
leads to the expectation $\delta \theta_r \sim \Theta$, i.e.,
the effective scalar increments contributing to the second moment
of scalar gradients are as strong as the rms of scalar fluctuations.
Indeed, this is in line with development of 
sharp cliff or fronts in the scalar field,
which have been routinely observed and investigated
in the literature \citep{KRS91, ShrSig00, celani2001, BCSY2021a}.

The observation that $h_2^* \to 0$ for second moment of scalar
gradient, has several important implications.
First, assuming boundedness of scalar increments (in the limit $r/L\to0$),
it follows that $h_2= 0$ also represents the smallest possible
value for the exponent \citep{Frisch95}.
For the velocity field, the second moment of velocity gradient
(corresponding to energy dissipation anomaly) 
corresponds to H\"older exponent $h \approx 1/3$,
with higher order moments coming from smaller values \citep{Frisch95}.
From boundedness of velocity increments, the expected $h_{\rm min} = 0$, 
with $\delta u_r \sim U$ 
\citep{Paladin87, BPBY2019, SreeniYakhot:2021, BP2022}.
However, for scalar gradients, 
$h_2^* \approx 0$ for second moment implies 
that even higher order moments must necessarily come from 
a critical exponent of zero. Thus, the sharp scalar cliffs
essentially control the scaling of all scalar gradient moments.
The results, formally presented in the next subsection, 
are indeed in perfect agreement with this expectation.

The second relates to the saturation of scaling 
exponents of scalar structure functions. 
If the minimum possible value of the exponent $h_2$
is zero, it is easy to see from Eq.~\eqref{eq:zetap},
that $\zeta_{0,p_2}$ must saturate 
for sufficiently large $p_2$, since higher moment
orders for structure functions also arise from
smaller values of critical exponents $h_2^*$.
Note that while $h_1$ does impact the scaling of
scalar gradient moments, which are related to mixed
velocity-scalar structure functions under the joint
multifractal framework, it does not play a direct role 
when only considering scalar structure functions. 
Several studies in past have suggested
saturation of scalar scaling exponents, 
with DNS results at high $\re$ also confirming
it \citep{ShrSig00, falkovich01, celani2001, gotoh2015power, KI18, BCSY2021b}. 
Thus, the existence of a scalar dissipation anomaly
also necessitates saturation of inertial-range scaling exponents of
scalar structure functions.

Another implication is related to the result in Eq.~\eqref{eq:yaglom_dh}, 
from which one can extract a fractal dimension corresponding to the 
mean-field. Taking $h_1^* = \tfrac{1}{3}$ for the velocity field,
the fractal dimension of 
leading to: 
\begin{align}
D^* = D(h_1^*, h_2^*) = 2\tfrac{1}{3}  + 2h_2^*  
\end{align}
At sufficiently high $\re$, when $h_2^* \approx 0$,
it follows that the effective fractal dimension 
of joint velocity-scalar field is $D^* \approx 2\tfrac{1}{3}$.
In contrast, energy dissipation anomaly 
corresponds to the result $h_1 = \tfrac{1}{3}$
with $D(h_1) \approx 3$. Thus, unlike the velocity field,
the fractal correspondence for scalar dissipation anomaly
is not space-filling. 
Several prior studies have studied the 
fractal properties of the scalar field alone
and reported a similar observations 
\citep{prasad1990_jfm, constantin91, grossmann1994, gauding2022}.
Though some care must be taken in making the comparisons, as 
our result is not for scalar field alone, but for
the joint fractal dimension of velocity and scalar field.

\subsection{Higher-order moments of scalar gradient}

Starting from Eq.~\eqref{eq:derv_n1n2},
and setting $n_1=0$ and $n_2 = n$, we can isolate
the scaling of scalar gradient moments as
\begin{align}
\langle 
(\partial_x \theta )^n \rangle \sim  
\left(\frac{\Theta}{L}\right)^{n} \ 
Re^{\rho_{0,n}}  \ Sc^{\sigma_n} \ . 
\label{eq:scgrad_scaling}
\end{align}
where $\rho_{0,n} = p - n$, with $p$ being the solution
to Eq.~\eqref{eq:zeta_t} and $\sigma_n = \tfrac{n}{2} (1-h_2^*)$,
with $h_2^*$ being the singularity exponent selected by the $n$-th order
gradient moment (and generally is expected to depend on $n$). 
Since $\rho_{0,2} =  1$, we can write an expression for the
scaling of central moments of scalar gradients
\begin{align}
\frac{\langle (\partial_x \theta )^n \rangle}
{\langle (\partial_x \theta )^2 \rangle^{n/2}} \sim  
Re^{\rho_{0,n} - n/2}  \ Sc^{\sigma_n - n \sigma_2/2}  \ . 
\label{eq:scg_n}
\end{align}
Thus, the joint framework predicts that 
scalar gradient moments are governed by two coupled effects. 
The $Re$-dependence is inherited from joint inertial-range
exponents of mixed velocity-scalar structure functions. On the other hand,
the $Sc$-dependence arises purely from sub-viscous scale 
amplification of scalar gradients in the regime
$\eta_\theta < r < \eta_u$.

\begin{figure}
\centering
\includegraphics[width=0.54\linewidth]{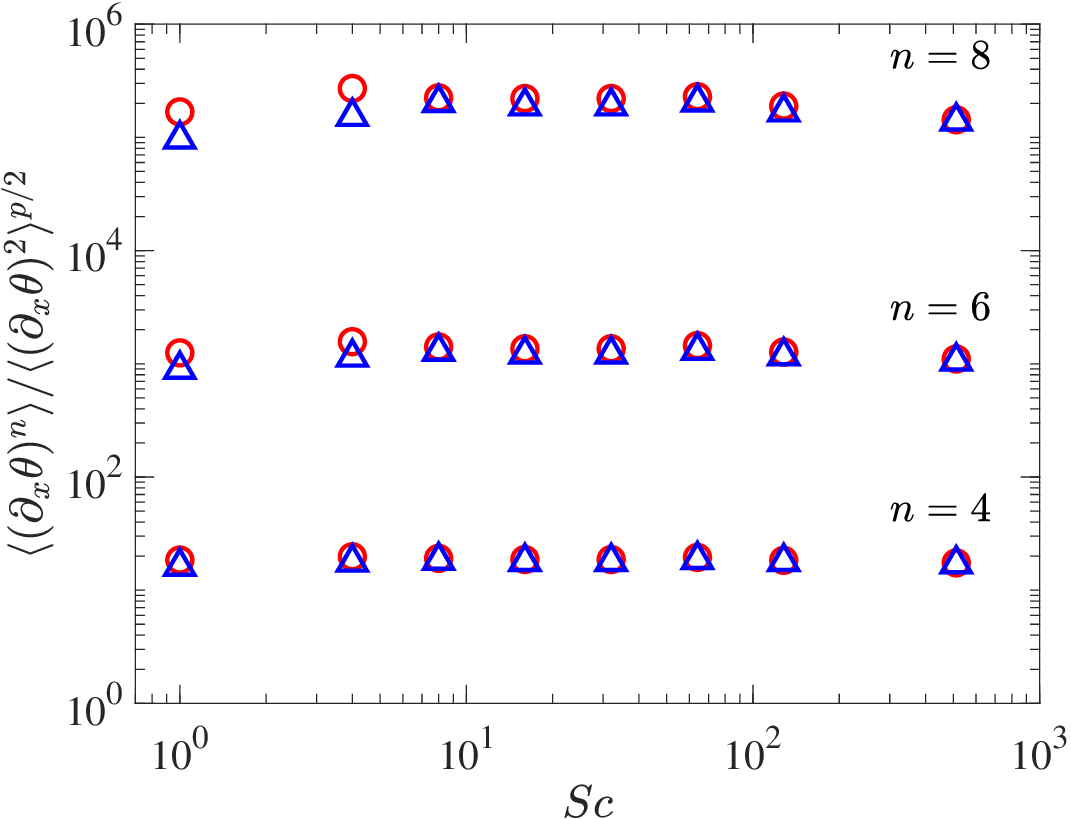}
\caption{
Even-order central moments of scalar gradient
components parallel (red circles) and perpendicular 
(blue triangles) to the direction of the imposed
mean-gradient, as a function of $Sc$, at $\re=140$. 
}
\label{fig:scmom_even}
\end{figure}

Based on our previous discussion
around $h_2^*$ in \S~ref{subsec:h2},
we now make the simplifying ansatz that the same 
$h_2^*$
contributes to scaling of all scalar gradient
moments. This leads to the expectation that 
$\sigma_n = \tfrac{n}{2} (1-h_2^*)=  \tfrac{n}{2} \sigma_2$,
which predicts that the central moments of scalar gradients
given by Eq.~\eqref{eq:scg_n} are independent of $Sc$.
This is indeed consistent with prior DNS results \citep{PK02, BCSY2021a},
and also with separate analyses of \citet{yasuda20, tang2023}
looking at other scalar statistics. 
Figure~\ref{fig:scmom_even} 
shows the scaling of even-order central moments 
as a function of $Sc$ for $\re=140$, as taken from 
\citet{BCSY2021a}. The different data points show
the gradient components parallel and perpendicular  
to the imposed mean-gradient; 
no discernible dependence on $Sc$ is observed 
for $Sc \gtrsim 8$.

\begin{figure}
\centering
\includegraphics[height=0.36\linewidth]{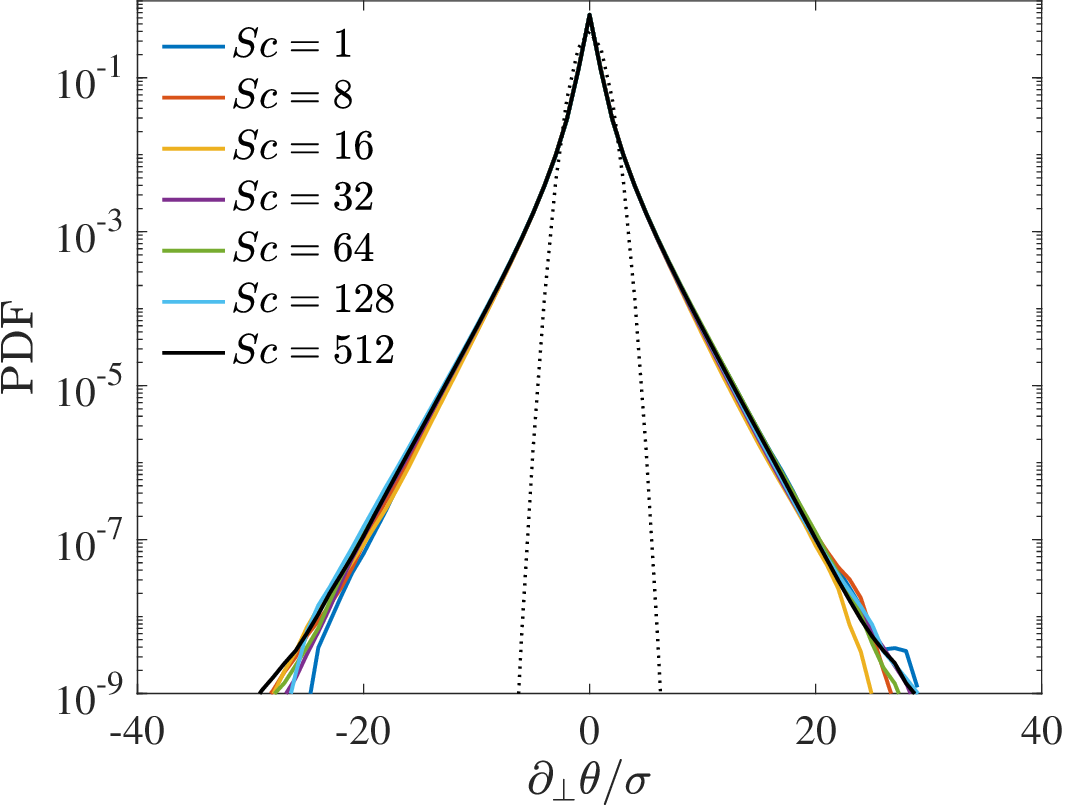} \ \ 
\includegraphics[height=0.36\linewidth]{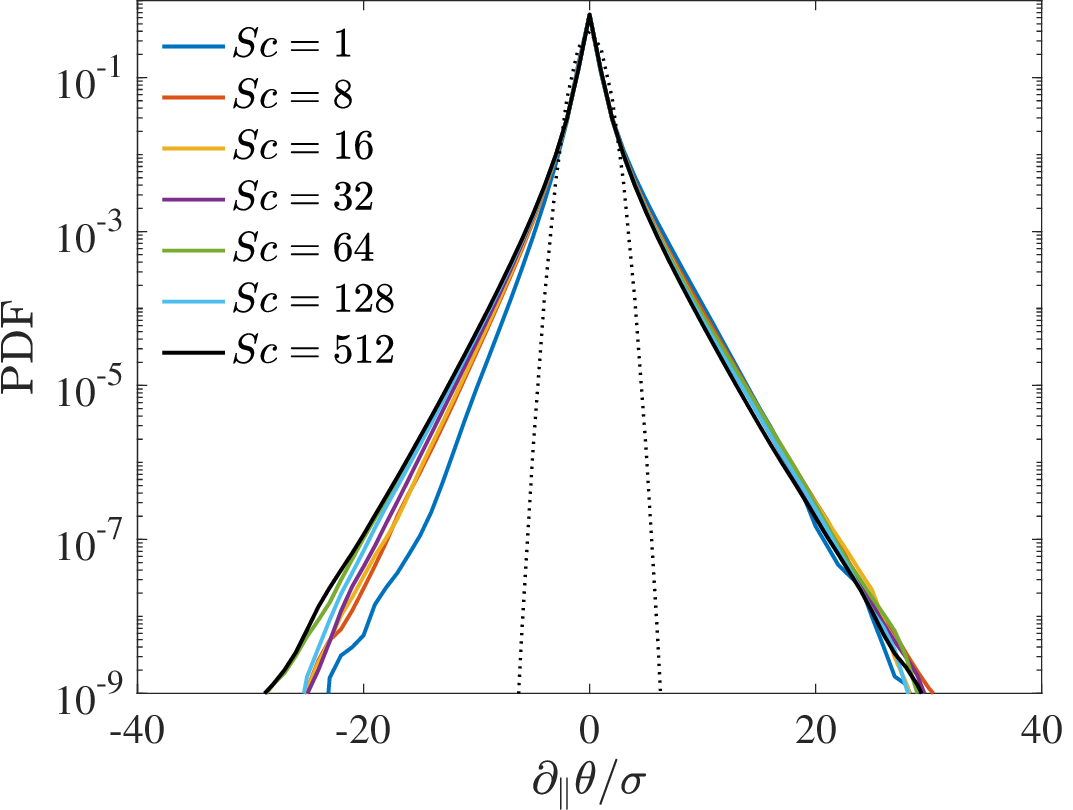}
\caption{
The standardized probability density functions (PDFs) of
scalar gradient components perpendicular (panel a)
and parallel (panel b) to the direction of the imposed
mean-gradient, at various $Sc$ for fixed $\re=140$. 
}
\label{fig:pdf_perp}
\end{figure}

\begin{figure}
\centering
\hspace{-0.2cm}
\includegraphics[height=0.36\linewidth]{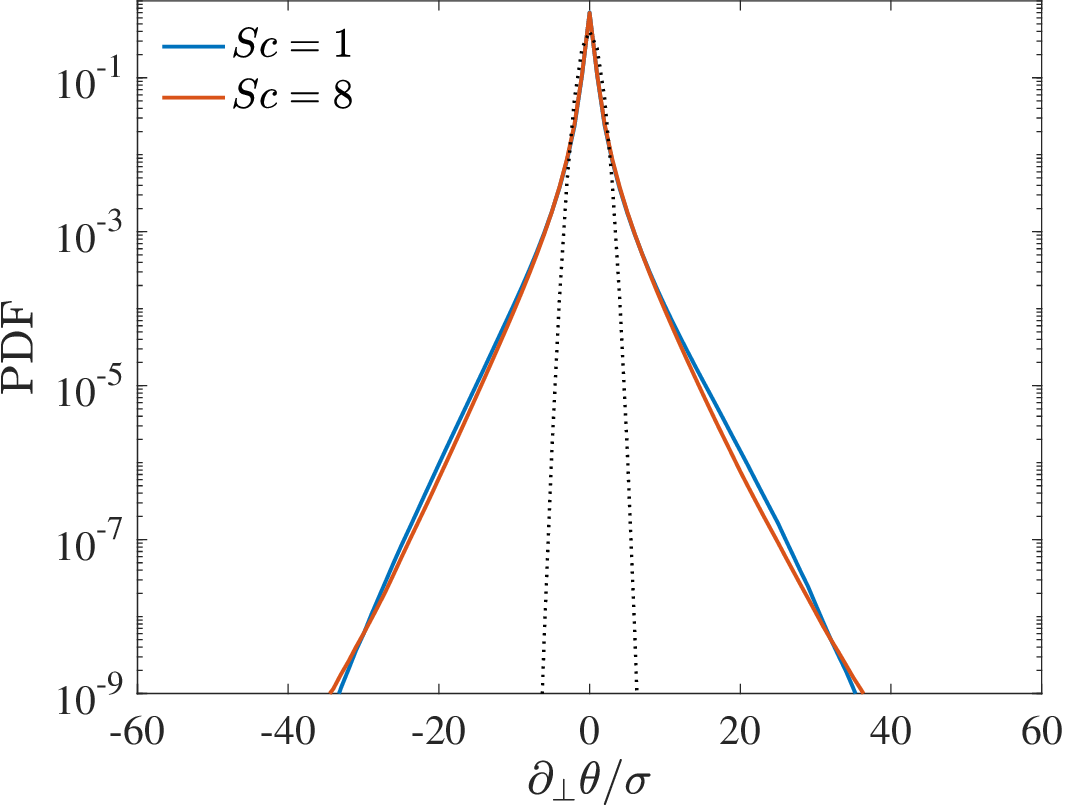} \ \ \ 
\includegraphics[height=0.36\linewidth]{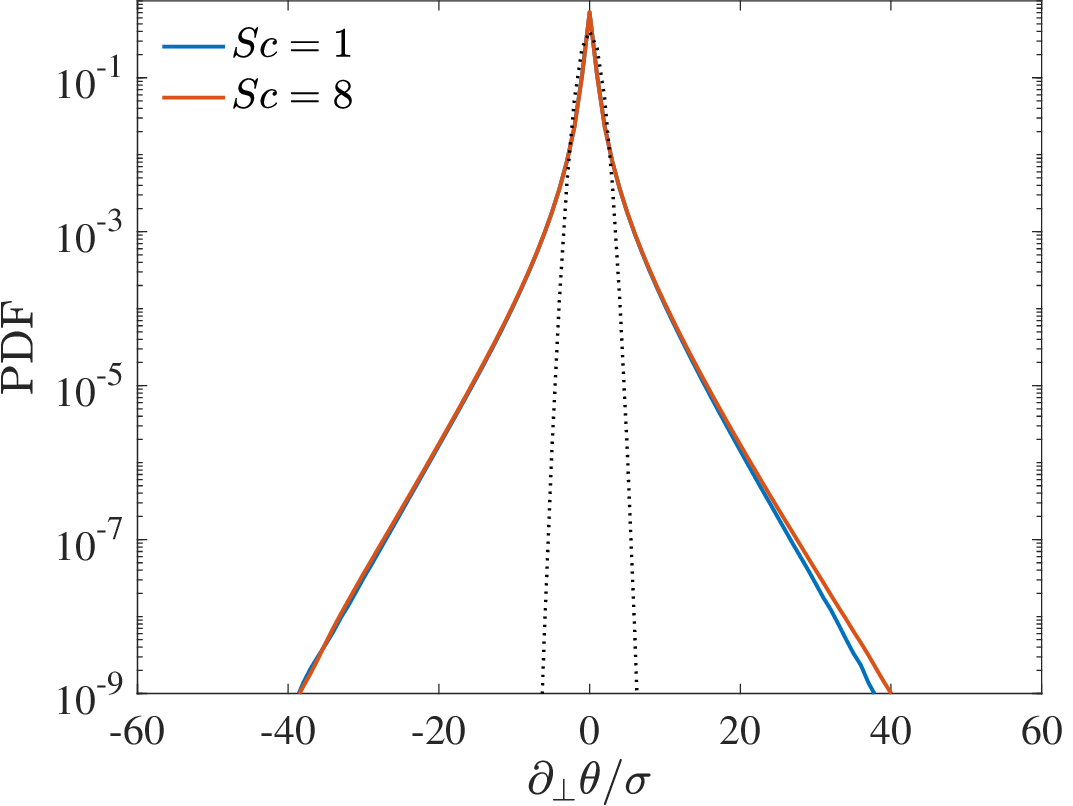}
\caption{
The standardized probability density functions (PDFs) of
scalar gradient component perpendicular to the direction of the imposed
mean-gradient for $Sc=1$ and $8$, at a) $\re=390$, and b) $\re=650$. 
}
\label{fig:pdf_re}
\end{figure}

The same conclusion can be examined more
directly by considering the standardized probability density functions (PDFs) 
of the scalar gradient. If all the central moments
are independent of $Sc$, we can expect a universal $Sc$-independent 
form of the standardized PDFs at any fixed $\re$. 
Figure~\ref{fig:pdf_perp}a shows the standardized PDFs
of scalar gradient components perpendicular 
to the imposed mean-gradient at different $Sc$ values for
$\re=140$. We can readily observe that the PDFs at different $Sc$ essentially 
collapse onto a single curve (outside of some very minor
statistical error for the most extreme events).
The corresponding results for $\re=390$ and $\re=650$ 
are shown in Fig.~\ref{fig:pdf_re}a and b, respectively.
Once again, we can observe that the curves for $Sc=1$ and $8$
are essentially coincident, supporting our earlier
prediction that central moments of scalar gradients, 
are independent of $Sc$.

The corresponding PDFs for the scalar gradient component parallel
to the mean gradient, at different $Sc$ values for
$\re=140$, are shown in Fig.~\ref{fig:pdf_perp}b.
In this case, we observe that the PDFs collapse 
for positive gradients, whereas for the negative gradients
exhibit a systematic $Sc$-dependence before seemingly 
approaching collapse at very high $Sc$. 
This behavior is expected as it is well-known that the 
parallel gradient component retains a measurable departure 
from local isotropy due to the imposed mean gradient, 
manifested in a finite positive skewness that decreases with 
increasing $Sc$ \citep{KRS91, PK02, BCSY2021a}. 
At sufficiently high $Sc$, the skewness approaches values 
close to zero, leading to nearly symmetric PDFs. However, note that 
this asymmetry has no significant effect on even-order moments, 
which are essentially identical for the parallel and perpendicular 
gradient components, as observed in Fig.~\ref{fig:scmom_even}.

It is worth emphasizing that the multifractal framework
implicitly assumes local isotropy, 
since the increments are characterized only by their 
separation distance and not by their orientation relative 
to the imposed mean scalar gradient.
The imposed mean gradient breaks this symmetry and gives rise to 
finite odd-order moments of the parallel scalar-gradient component. 
Consequently, the present framework is not intended to describe 
the skewness or other anisotropic odd-order statistics of 
the parallel gradient.
These quantities instead require a description that 
explicitly accounts for the ramp-cliff organization of the scalar 
field induced by the mean gradient \citep{krs77, BCSY2021a}. 
For this purpose, we refer the reader to the recent ramp-cliff model developed 
in \citet{BCSY2021a}, which was shown to capture the observed $Sc$-dependence 
of odd central moments with remarkable accuracy.

As a final note, we draw attention to the Reynolds number
dependence of the standardized scalar-gradient PDFs
in Fig.~\ref{fig:pdf_perp}a and Fig.~\ref{fig:pdf_re}. 
While the PDFs at a fixed $\re$ are essentially independent
of $Sc$, their tails broaden systematically 
as $\re$ increases. That is, extreme scalar-gradients
become more frequent as the $\re$ increases, analogous
to velocity-gradient intermittency \citep{Ishihara09, buaria_jfmp}. 
The Reynolds number scaling of gradient moments
is contained in
Eq.~\eqref{eq:scg_n} through the exponent $\rho_{0,n}$.
Unlike the $Sc$ dependence, however, this exponent is non-trivial;
it depends on the joint multifractal spectrum, and equivalently,
on the scaling exponents of mixed velocity-scalar structure functions.
A detailed analysis of this connection 
will be explored in future work. Nevertheless, the present
results suggest an important distinction: scalar gradient
intermittency is essentially a Reynolds-number effect,
similar to velocity-gradient intermittency. 
Whereas increasing $Sc$ simply rescales the gradient
magnitude through the second moment, as prescribed by scalar
dissipation anomaly.

\section{Summary and conclusions}

In this work, we have investigated scalar dissipation anomaly and scalar-gradient
scaling in passive scalar turbulence at high Reynolds and
Schmidt numbers. The central objective is to understand 
the behavior of mean scalar dissipation rate
in the joint limit of large $\re$ and $Sc$, 
and how it relates
to the scaling of scalar gradients. 
To this end, we developed a joint multifractal
framework for velocity and scalar statistics
and tested its predictions using
high-resolution data from direct numerical
simulations (DNS).
The framework is analogous to that recently developed 
for longitudinal and transverse velocity increments \citep{buaria2026}, 
but here it is formulated in terms of longitudinal velocity 
increments and scalar increments. This allows Yaglom's law 
to be used as an exact constraint on the joint scaling. 
Extending the framework to scalar gradients for $Sc>1$ requires 
accounting for the fact that the scalar cutoff scale lies below the viscous 
cutoff of the velocity field. This leads naturally to a 
fluctuating Batchelor scale and yields predictions for 
scalar-gradient moments in terms of the joint 
velocity-scalar multifractal spectrum.
The framework naturally predicts
Reynolds number independence of the normalized
mean scalar dissipation rate, but a residual $Sc^{-h_2^*}$ 
dependence emerges, where $h_2^*$ is the critical
scalar H\"older exponent contributing to the scaling. 
Thus, a strict scalar dissipation anomaly
with respect to both $\re$ and $Sc$ is realized only when
$h_2^* \to 0$, corresponding to scalar increments
as strong as the rms of scalar fluctuations, corresponding
to sharp cliffs or fronts that are  
characteristic of turbulent mixing.

Analysis of the DNS data confirms this prediction.
For $Sc=1$, the normalized mean scalar dissipation rate
asymptotes to a constant for $\re \gtrsim 140$,
consistent with previous studies \citep{Donzis05, BCSY2021b} 
and extended here to $\re=1000$. 
For $Sc=8$, the scalar dissipation is suppressed 
at low Reynolds numbers, but progressively
recovers the same asymptotic value at $\re \gtrsim 650$.
We empirically observe that the critical exponent follows
$h_2^* \approx 29.5/\re$,
so that $h_2^* \to 0$ and a true anomaly is recovered 
in the large-$\re$ limit. 
Since $h_2^*$ is small at high $\re$, the power-law
dependence is identical to a
$\log Sc/\re$ correction for the reciprocal
of normalized scalar dissipation, in agreement
with previous predictions by \citet{borgas2004high, Donzis05}. 
The simple conclusion is therefore that scalar dissipation 
anomaly applies at any $Sc$, provided the Reynolds number is 
sufficiently high, with larger $Sc$ requiring larger $\re$ to reach 
the asymptotic state.

The result $h_2^* \to 0$ carries several further implications.
Since $h_2 =0$ is also the smallest admissible scalar
exponents, the sharp fronts that fix the second moment
also control the $Sc$-dependence of all higher-order
moments of scalar gradient, and additionally
imply saturation of 
inertial-range exponents of scalar structure functions. 
The same condition fixes the joint velocity-scalar
fractal dimension to $D^* \approx 7/3$, which suggests that
the structures responsible for scalar dissipation 
occupy a non-space filling set -- unlike
the space-filling $D^* \approx 3$ associated with energy
dissipation anomaly. 
Since the higher order moments are fixed by the second moment,
it directly follows that standardized scalar gradient statistics
are independent of $Sc$ at fixed $\re$. DNS data
indeed confirm this through both the $Sc$-independence
of central moments, and the collapse of standardized
PDFs of scalar gradients across $Sc$ at each $\re$. 
add one sentence about mean gradient part.
The scalar-gradient component parallel to the 
imposed mean gradient retains finite odd-order anisotropy,
connected to underlying ramp-cliff structures
\citep{BCSY2021a}, 
but this does not affect the even-order moments considered 
in the present locally isotropic multifractal framework.

In contrast, scalar gradients exhibit a strong Reynolds
number dependence, with PDF tails broadening 
as $\re$ increases, in close analogy with the intermittency
of velocity gradients.
In the present joint framework, the corresponding 
Reynolds number exponents of scalar gradient moments 
are tied to mixed velocity-scalar
structure functions, and therefore the full joint
multifractal spectrum. This is analogous to the
scaling of transverse velocity gradients, which 
can be related to mixed longitudinal-transverse 
structure function exponents \citep{buaria2026}. 
Thus, increasing $Sc$ primarily shifts the scalar cutoff
to smaller sub-viscous scales and rescales the gradient magnitude
through the second moment. Whereas increasing $\re$ 
changes the intermittent structure of the scalar gradient
field itself. Determining this Reynolds-number dependence 
quantitatively, together with the associated mixed velocity-scalar 
scaling exponents, is left for future work.



\begin{bmhead}[Acknowledgements.]
We gratefully acknowledge the Gauss Centre for Supercomputing 
e.V. (www.gauss-center.eu) for providing time on the supercomputer
JUWELS at J\"ulich Supercomputing Centre (JSC).
We also acknowledge the Texas Advanced Computing Center (TACC) 
at UT Austin (www.tacc.utexas.edu) for providing computational 
resources that have contributed to the research results reported 
within this paper.  
The high Schmidt number simulations
using the hybrid approach were performed together 
with Matthew P. Clay and P. K. Yeung using computational resources at 
the Oak Ridge Leadership Computing Facility (OLCF), under 2017
and 2018 INCITE Awards.
\end{bmhead}


\begin{bmhead}[Declaration of interests.]
The authors report no conflict of interest.
\end{bmhead}

\begin{bmhead}[Data availability statement.]
The data that support the findings of this study are 
available from the corresponding author 
upon reasonable request.
\end{bmhead}

\begin{bmhead}[Author ORCID.]
Dhawal Buaria, https://orcid.org/0000-0001-9167-1712 
\end{bmhead}

\bibliographystyle{jfm}


\end{document}